\documentclass[12pt,a4paper,twoside]{article}
\usepackage[T1]{fontenc}
\usepackage[applemac]{inputenc}
\usepackage{amsmath,amsfonts,amssymb,amsthm,mathabx}
\usepackage{cite}
\usepackage{hyperref}
\usepackage{dsfont}
\usepackage{graphicx}
\usepackage{color}
 
\usepackage{pdfsync}

\newcommand{\be}{\begin{equation}}
\newcommand{\ee}{\end{equation}}
\newcommand{\bea}{\begin{eqnarray}}
\newcommand{\eea}{\end{eqnarray}}

\newcommand{\Ad}{\mathrm{Ad}}

\def\Item$#1${\item $\displaystyle#1$
   \hfill\refstepcounter{equation}(\theequation)}

\def\d_Vphi{\mathrm{d}_V\hspace{-0.06em}\phi}
\def\d_Vphibar{\mathrm{d}_V\hspace{-0.06em}\bar\phi}
\def\d_Vxi{\mathrm{d}_V\hspace{-0.06em}\xi}

\def\be{\begin{eqnarray}}
\def\ee{\end{eqnarray}}
\def\beann{\begin{eqnarray*}}
\def\eeann{\end{eqnarray*}}
\def\beq{\begin{equation}}
\def\eeq{\end{equation}}
\def\ba{\begin{array}}
\def\ea{\end{array}}
\def\ben{\begin{enumerate}}
\def\een{\end{enumerate}}
\def\bea{\begin{eqnarray}}
\def\eea{\end{eqnarray}}

\def\5{\bar }
\def\6{\partial }
\def\7{\hat }
\def\4{\tilde }
\advance\textheight\topskip
\renewcommand{\tilde}{\widetilde}
\renewcommand{\hat}{\widehat}

\newcommand{\bref}[1]{\textbf{\ref{#1}}}

\newcommand{\vp}{\varphi}

\renewcommand{\d}{\partial}

\newcommand{\binner}[2]{%
  {\langle}\kern-4.15pt{\langle}#1{,}\,#2{\rangle}\kern-4.15pt{\rangle}}

\newcommand{\half}{\frac{1}{2}}

\newcommand{\ffrac}[2]{\raisebox{.5pt}%
  {\footnotesize$\displaystyle\frac{#1}{#2}$}\kern1pt}

\def\cA{\mathcal{A}}

\def\cH{\mathcal{H}}

\def\cJ{\mathcal{J}}

\def\cM{\mathcal{M}}
\def\cN{\mathcal{N}}

\def\cP{\mathcal{P}}

\def\cS{\mathcal{S}}

\numberwithin{equation}{section} \makeatletter
\@addtoreset{equation}{section}

\DeclareFontFamily{OT1}{rsfs}{} \DeclareFontShape{OT1}{rsfs}{m}{n}{
<-7> rsfs5 <7-10> rsfs7 <10-> rsfs10}{}
\DeclareMathAlphabet{\mycal}{OT1}{rsfs}{m}{n}

\begin{document}

\title{Geometric actions for three-dimensional gravity}

\author{Glenn Barnich, Hern\'an A. Gonz\'alez, Patricio Salgado-Rebolledo}

\date{}

\def\mytitle{Geometric actions for three-dimensional gravity}

\pagestyle{myheadings} \markboth{\textsc{\small G.~Barnich,
    H.A.~Gonz\'alez, P.~Salgado-Rebolledo}}{%
  \textsc{\small Geometric actions for 3d gravity}}

\addtolength{\headsep}{4pt}

\begin{centering}

  \vspace{1cm}

  \textbf{\Large{\mytitle}}


  \vspace{1.5cm}

  {\large G.~Barnich$^1$, H.~A.~Gonz\'alez$^2$ and
    P.~Salgado-Rebolledo$^{3}$ }

\vspace{.5cm}

\begin{minipage}{.9\textwidth}\small \it  \begin{center}
    $^1$Physique Th\'eorique et Math\'ematique \\ Universit\'e Libre
    de Bruxelles and International Solvay Institutes \\ Campus Plaine
    C.P. 231, B-1050 Bruxelles, Belgium
 \end{center}
\end{minipage}

\vspace{.5cm}

\begin{minipage}{.9\textwidth}\small \it  \begin{center}
$^2$Institute for Theoretical Physics\\
Vienna University of Technology\\ 
Wiedner Hauptstr. 8-10/136, A-1040, Vienna, Austria
 \end{center}
\end{minipage}

\vspace{.5cm}

\begin{minipage}{.9\textwidth}\small \it  \begin{center}
   $^3$Facultad de Ingenier\'ia y Ciencias and UAI Physics Center\\
   Universidad Adolfo Iba\~nez\\
   Avda. Diagonal las Torres 2640, Pe\~nalol\'en, Santiago, Chile 
 \end{center}
\end{minipage}

\end{centering}

\vspace{1cm}

\begin{center}
  \begin{minipage}{.9\textwidth}
    \textsc{Abstract}. The solution space of three-dimensional
    asymptotically anti-de Sitter or flat Einstein gravity is given by
    the coadjoint representation of two copies of the Virasoro group
    in the former and the centrally extended BMS$_3$ group in the
    latter case. Dynamical actions that control these solution spaces
    are usually constructed by starting from the Chern-Simons
    formulation and imposing all boundary conditions. In this
    note, an alternative route is followed. We study in detail how to
    derive these actions from a group-theoretical viewpoint by
    constructing geometric actions for each of the coadjoint orbits,
    including the appropriate Hamiltonians. We briefly sketch relevant
    generalizations and potential applications beyond three-dimensional 
    gravity.
  \end{minipage}
\end{center}

\vspace{1cm}
\thispagestyle{empty}
\newpage

\begin{small}
{\addtolength{\parskip}{-2pt}
 \tableofcontents}
\end{small}
\thispagestyle{empty}
\newpage

\section{Introduction}
\label{sec:introduction}

Even though three-dimensional general relativity does not admit local
degrees of freedom, there is both an infinite dimensional symmetry
structure \cite{Brown:1986nw} and rich dynamics in two dimensions
\cite{Coussaert:1995zp} when allowing for non-trivial boundary
conditions. The standard ``bulk'' approach to derive this dynamics
starts from the Chern-Simons formulation
\cite{Achucarro:1987vz,Witten:1988hc}: when taking into account the
boundary conditions and the associated surface terms
\cite{Regge:1974zd}, this leads one in a first step to a
Wess-Zumino-Witten theory along the lines of
\cite{Witten:1988hf,Moore:1989yh,Elitzur:1989nr}, and in a second step
to Liouville theory through a Hamiltonian reduction
\cite{Forgacs:1989ac,Bershadsky:1989mf,Alekseev:1988ce}. In this
approach, the Hamiltonian of the dual theory is inherited from the
surface term associated with time translations. Alternatively, the
dual theory may be contructed through holographic renormalization in
the context of the AdS$_3$/CFT$_2$ correspondence
\cite{Skenderis:1999nb}.

In this paper, we follow a different method to construct dual
two-dimensional action principles for the gauge fixed solution spaces
of three-dimensional gravity. Indeed, both in the asymptotically AdS
and flat cases, the solution space coincides with the centrally
extended coadjoint representation, at fixed values of the central
charges, of the asymptotic symmetry groups, viz.~two copies of the
Virasoro group
\cite{Brown:1986nw,Banados:1998gg,Skenderis:1999nb,NavarroSalas:1999sr,%
  Nakatsu:1999wt,Garbarz:2014kaa,Barnich:2014zoa} and the centrally
extended $\widehat{\rm BMS}_3$ group
\cite{Ashtekar:1996cd,Barnich:2006av,Barnich:2010eb,Barnich:2011ct,%
  Barnich:2012rz,Barnich:2014kra,Barnich:2015uva}, respectively (see
also \cite{Raeymaekers:2014kea,Garbarz:2015lua,%
  Campoleoni:2016vsh,Oblak:2016eij,Oblak:2017ect} for recent related
considerations). More precisely, the general solution to the Einstein
equations with AdS asymptotics is given by
\begin{equation}
  \label{eq:1}
  ds^2=\frac{l^2}{r^2}dr^2-(rdx^+-\frac{8\pi Gl}{r}b^-dx^-)
  (rdx^--\frac{8\pi Gl}{r}b^+dx^+), 
\end{equation}
with $x^\pm=\frac{t}{l}\pm\varphi$ and the arbitrary $2\pi$-periodic
functions $b^\pm(x^\pm)$ transforming as
\begin{equation}
  \label{eq:3}
  \tilde b^\pm=(\d_\pm f^\pm)^2 b^\pm\circ f^\pm-c^\pm
  S_{x^{\pm}}[f^\pm], \quad c^\pm=\frac{3l}{2G}, 
\end{equation}
under the conformal transformations $x^\pm\to f^\pm(x^\pm)$,
$f^\pm(x^\pm+2\pi)=f^\pm(x^\pm)\pm 2\pi$ of the cylinder
at infinity, with the Schwarzian derivative given by
\begin{equation}
  \label{eq:5}
  S_x[f]=\frac{1}{24\pi}\left[\d^2_x(\log \d_xf)-\half (\d_x(\log \d_x f))^2\right]. 
\end{equation}
This coincides with the coadjoint representation of the two copies of
the Virasoro group for fixed values of the central charges.  For
asymptotically flat spacetimes, one finds instead
\begin{equation}
  \label{eq:2}
  ds^2=2[8\pi G p du-dr +8\pi G(j+up')d\varphi]du+r^2d\varphi^2, 
\end{equation}
where the arbitrary $2\pi$-periodic functions $p=p(\varphi),j=j(\varphi)$
transform as
\begin{equation}
  \label{eq:4}
  \begin{split}
  \tilde p =(f')^2 p \circ f-c_2
  S_{\vp}[f], \quad c_2=\frac{3}{G},\\
  \tilde j=(f')^2 [j+\alpha p' +2\alpha'
  p-\frac{c_2}{24\pi}\alpha''']\circ f-c_1 S_{\vp}[f], \quad c_1=0, 
\end{split}
\end{equation}
under the ${\rm BMS}_3$ transformations
$(u,\varphi)\to (f'u+\alpha(f(\varphi)),f(\varphi))$, with
$f(\varphi+2\pi)=f(\varphi)+2\pi$ and
$\alpha(\varphi+2\pi)=\alpha(\varphi)$. In turn, this coincides wih
the coadjoint representation of the centrally extended ${\rm BMS}_3$
group for fixed values of the central charges.

As a consequence, the gravitational solution space admits a partition
into coadjoint orbits. For any group $G$, the individual orbits are
homogeneous symplectic spaces $G/{\cH}$ (see
e.g.~\cite{Kostant1970,Souriau1970,kirillov1976elements,Kirillov2004}
and original references therein). The aspect we will exploit here is
that one can associate to each of these orbits, in a canonical way,
geometric actions which admit $G$ as a global and ${\cH}$ as a gauge
symmetry group \cite{Alekseev:1988vx}. What is fixed in these actions
is the kinetic term determined by the symplectic structure on the
coadjoint orbit. If the aim is to construct $G$-invariant dynamical
systems, one may choose a suitable Hamiltonian defined on the
coadjoint orbit that respects $G$-invariance (see also
e.g.~\cite{guieu2007,Khesin2009} for reviews).

When this method is applied to three-dimensional gravity, we will get
finer actions than those of
\cite{Coussaert:1995zp,Henneaux:1999ib,Barnich:2012rz,Barnich:2013yka},
precisely adapted to the individual orbits. From the bulk point of
view, they take additional information on non-trivial holonomies into
account. In the anti-de Sitter case for example, one finds an
intriguing connection between $3d$ and $2d$ gravity in the sense that
the geometric actions for each copy of the Virasoro group differ from
the action for two-dimensional gravity by taking into account a more
general covector, for which not only the central charge differs from
zero \cite{Alekseev:1988ce,Rai:1989js,doi:10.1142/S0217751X90001690,%
  Aratyn:1990dj,Nissimov:2000gw}. In the flat case, this approach
allows us to 
construct novel $\widehat{\rm BMS}_3$ invariant actions (see also
\cite{Rebolledo2015}). Note that the Hamiltonian can be fixed by
reverting to the bulk approach described above in order to identify a
suitable generator.

Another interesting aspect of this approach to two-dimensional
conformal or BMS$_3$ invariant actions is that, exactly like in the
case of loop groups and the associated Wess-Zumino-Witten theories,
they can also be interpreted as one-dimensional particle/world-line
actions associated to infinite-dimensional groups. The spatial
dimension is hidden or emergent, depending on whether one uses a
Fourier expansion for the Lie algebra generators and their duals with
associated infinite mode sums or an inner product with an explicit
integration over the circle. For instance, from the two-dimensional
point of view, the deformation by a Wess-Zumino-Witten term involves a
3-cocycle of the Lie algebra of the group $G$, whereas in the
worldline approach, this deformation comes from a related $2$-cocycle
on the Lie algebra of the loop group of $G$.

The structure of the paper is as follows: in the next section, we
review the construction of geometric actions, with a special
emphasis on how to include Hamiltonians that preserve
$G$-invariance. We continue with a discussion of central extensions
and the well-known examples of the Kac-Moody and Virasoro group
relevant for three-dimensional asymptotically anti de-Sitter
gravity. We then move to semi-direct product groups in order to be
able to treat three-dimensional flat gravity. In this section, we
provide novel $\mathfrak{iso}(2,1)$ WZW models and BMS$_3$ invariant
field theories in two dimensions. In the last section, we relate our
considerations to recent developments in the field and discuss future
prospects, both for three and four dimensional gravity.

\section{Review of geometric actions}
\label{I}

\subsection{Kinetic term}
\label{sec:kinetic-term}

The adjoint action of a Lie group $G$ on its Lie algebra $\mathfrak{g}$
is defined as the differential of the automorphism $h\mapsto g h g^{-1}$
at the identity 
\begin{equation}
{\rm Ad}_{g}X=\left.\frac{d}{ds}\left(g
    h\left(s\right)g^{-1}\right)\right\vert _{s=0},\label{eq:adg} 
\end{equation}
where $X=\left.\frac{dh\left(s\right)}{ds}\right\vert _{s=0}\in{\rm \mathfrak{g}}$.
The coadjoint action of $G$ on the dual space ${\rm \mathfrak{g}}^{*}$
of $\mathfrak{g}$ is defined as
\begin{equation}
  \left\langle {\rm Ad}_{g^{-1}}^{*}b,X\right\rangle
  =\left\langle b,{\rm Ad}_{g}X\right\rangle, \label{eq:ad*g}
\end{equation}
where $b\in\mathfrak{g}^{*}$ and
$\left\langle\cdot ,\cdot\right\rangle $ is the pairing between
$\mathfrak{g}$ and ${\rm \mathfrak{g}}^{*}$. For a fixed element
$b_{0}$ of ${\rm \mathfrak{g}}^{*}$, this action generates a coadjoint
orbit $O_{b_{0}}$, the set of elements $b\in \mathfrak g^*$ such that
\begin{equation}
b ={\rm Ad}^{*}_{g^{-1}}b_{0}\label{eq:7},
\end{equation}
which is a manifold
isomorphic to $G/{\cH}_{b_{0}}$, with ${\cH}_{b_{0}}$ the isotropy
group of $b_{0}$ under the coadjoint action,  i.e., the subgroup of
elements $h\in G$ satisfying ${\rm Ad}_{h}^{*} b_0 =b_0$. 

Coadjoint orbits are particulary interesting as they are symplectic
manifolds. The Kirillov-Kostant symplectic form is the pull-back to a
coadjoint orbit $O_{b_{0}}$ of the pre-symplectic form on $G$ given by
\begin{equation}
  \Omega= \frac{1}{2}\left\langle b, {\rm ad}_\theta \theta
  \right\rangle, \label{eq:sf} 
\end{equation}
where $b={\rm Ad}^{*}_{g^{-1}}b_{0}$ is a point on the orbit, $\theta$
is the left invariant Maurer-Cartan form satisfying
\begin{equation}
d\theta=-\frac{1}{2}{\rm ad}_{\theta}\theta,\label{eq:mceq}
\end{equation}
and ${\rm ad}$ denotes the adjoint action of $\mathfrak{g}$ on
itself. As $\Omega$ is closed, it is locally exact. In fact,
\begin{equation}
\label{KK}
\Omega=d a, \quad a=\left\langle b,\theta\right\rangle,
\end{equation}
and therefore, a geometric action $I_{\rm G}[g;b_0]$ can be defined on the phase
space through
\begin{equation}
I_{\rm G}[g;b_0]=\int_{\gamma} a,
\label{eq:ga}
\end{equation}
where $\gamma$ is a path on the coadjoint orbit $O_{b_{0}}$.

In particular, for finite dimensional matrix groups, a local solution
to \eqref{eq:mceq} is simply $\theta=g^{-1}dg$ and the pre-symplectic
potential becomes $a=\left\langle b_{0}, dg g^{-1}\right\rangle$. The
first order Euler-Lagrange equations of motion are equivalent to
setting to zero the one-forms $i_V\Omega$ where $V$ is the vector
field associated to $\frac{dg}{dt}$ and $t$ parametrizes the path
$\gamma$.  Note also that one may (trivially) write this action in
Wess-Zumino-Witten form,
\begin{equation}
  \label{eq:16}
  I_{\rm G}[g;b_0]=\int_\Sigma \Omega. 
\end{equation}
The assumptions here are that fields are extended to $\Sigma$ and that
$\gamma$ is part of the boundary of $\Sigma$, with suitable conditions
on fields and their derivatives such that all other boundary terms
vanish.  

Consider the vector fields associated to one parameter families of
right translations generated by $X$,
$v^R_X=\frac{d}{ds}(gh_R(s))|_{s=0}$ with $\frac{d}{ds}
h_R(s)|_{s=0}=X$. These vector fields are the left invariant vector
fields that reduce to $X$ at the identity. For all $X\in \mathfrak g$
that are constant along the path, they define global symmetries of
$I_{\rm G}[g;b_0]$. This follows from
\begin{equation}
i_{v^R_X}\Omega=dQ_X,\quad Q_X=-\langle b,X\rangle, \label{eq:6}
\end{equation}
and $\pounds_{v^R_X}
a=i_{v^R_X}\Omega+d\langle b,X\rangle$. The associated Noether charge
is $Q_X$.

Elements $\epsilon$ of ${\mathfrak h}_{b_0}$, the ``little'' algebra
associated to ${\cH}_{b_0}$, are defined by
${\rm ad}^*_\epsilon b_0=0$. Let $v^L_\epsilon$ be the vector fields
associated to one parameter families of left translations by elements
$h_L(s)$ of ${\cH}_{b_0}$ and generated by $\epsilon$,
$v^L_\epsilon=\frac{d}{ds}(h_L(s)g)|_{s=0}$ with
$\frac{d}{ds} h_L(s)|_{s=0}=\epsilon$. These vector fields are the
right invariant vector fields that reduce to $\epsilon$ at the
identity. Let us now assume furthermore that $\epsilon=\epsilon(t)$
depends a priori arbitrarily on $t$. It then follows from
$\pounds_{v^L_\epsilon}a=i_{v^L_\epsilon}\Omega+d\langle
b_0,\epsilon\rangle$ and
\begin{equation}
  \label{eq:8}
  i_{v^L_\epsilon}\Omega=0,
\end{equation}
that these transformations define gauge symmetries. More precisely,
the action $I_{\rm G}[g;b_0]$ is gauge invariant provided that $\epsilon(t)$
vanishes at the end points of $\gamma$. Note that $Q_X$
is gauge invariant since 
\begin{equation}
  \pounds_{v^L_\epsilon}
  Q_X=0\label{eq:9},
\end{equation}
by using \eqref{eq:6} and \eqref{eq:8}. Note also that the Noether
charges form a representation under the action of the global
symmetries,
\begin{equation}
  \label{eq:10}
  \pounds_{v^R_{X_1}} Q_{X_2}=Q_{[X_1,X_2]}. 
\end{equation}

\subsection{Symmetric Hamiltonians and deformations}
\label{sec:hamiltonians}

One can always add to the geometric action \eqref{eq:ga} a gauge
invariant function on the orbit playing the role of a Hamiltonian. The
inclusion of such a Hamiltonian generically breaks some of the global
symmetries of $I_{\rm G}[g;b_0]$. In the spirit of effective theory, one is
interested in Hamiltonians or other deformations that preserve global
symmetries.

(i) One possibility that preserves all global symmetries is to consider an
extended action, where one of the Noether charges plays the role of
the Hamiltonian, $H_X=Q_X$, so that the first order action becomes
\begin{equation}
\label{Hv}
I_{\rm G}[g;b_0,H_X]=\int_{\gamma}(a-H_X dt), 
\end{equation}
for some constant $X\in \mathfrak g$. The equations of motion are then
equivalent to the vanishing of 
\begin{equation}
  \label{eq:20}
  -i_V\Omega-d H_X=-(i_V+i_{v^R_X})\Omega. 
\end{equation}
By construction, this action is gauge invariant under the same
assumptions as before. It is also invariant under the global
transformations associated to $v^R_{X'}$. Indeed, by using \eqref{eq:10}, one finds that this is the case
for instance when $X'(t)$ depends explicitly on time with an evolution
determined by
\begin{equation}
  \label{eq:12}
  \frac{d X'}{dt}={\rm ad}_XX'. 
\end{equation}

(ii) Another possibility uses an invariant symmetric tensor on $\mathfrak
g^*$. Let $e_a$ denote the elements of a basis of $\mathfrak{g}$ and
$e^a$ the elements of the dual basis. The Noether charges assocated to
$e_a$ are $Q_a=- \langle b, e_a \rangle$. If $k^{a_1\dots a_m}$ denote
the components of the tensor, the Hamiltonian can be chosen to be,  
\begin{equation}
H_{k}=\frac{1}{m!} k^{a_1\dots a_m} Q_{a_1}\dots Q_{a_m}. 
\label{eq:13}
\end{equation}
From \eqref{eq:8} and \eqref{eq:10}, it follows that Lie derivatives
with respect to ${v^L_\epsilon}$ and ${v^R_{X'}}$ annihilate
$H_{k}$. Thus, geometric actions supplemented by $H_{k}$ also preserve
gauge and global symmetries, in this case with time independent
$X'$. An Hamiltonian quadratic in the Noether charges may be
constructed for instance for semi-simple Lie algebras by using the
inverse of the Killing form.

(iii) Another deformation of $I_{\rm G}[g,b_0]$ in
\eqref{eq:ga} that changes the kinetic term is given by
\begin{equation}
  \label{eq:23}
  I_{\rm G}[g;b_0,c]=I_{\rm G}[g,b_0]+\int_\Sigma \Omega_\omega,\quad
  \Omega_\omega=-\half c \omega({\rm Ad}_g
  \theta,{\rm Ad}_g \theta)
\end{equation}
with $\omega$ a Lie algebra 2-cocyle. 
Such a deformation is trivial in
the sense that it can be absorbed by a redefinition of $b_0$ if
$\omega$ is a coboundary. Non-trivial deformations are thus
characterized by $[\omega]\in H^2(\mathfrak g,\mathbb R)$.

These deformations modify the gauge symmetries: the requirement
$i_{v^L_\epsilon}\hat \Omega=0$ where
$\hat\Omega=\Omega+\Omega_\omega$ now restricts $\epsilon$ to solve
\begin{equation}
  \label{eq:21}
  {\rm ad^*}_\epsilon b_0-c s(\epsilon)=0, 
\end{equation}
with
\begin{equation}
\langle s(X),Y\rangle=-\omega(X,Y).\label{eq:18}
\end{equation}

Concerning global symmetries, the cocycle condition for $\omega$
implies that $di_{v^R_X}\Omega_\omega=0$. When taking into account
that
$i_{v^R_X}\Omega_\omega=-c\omega({\rm Ad}_gX,{\rm
  Ad}_g\theta)=c\langle {\rm Ad}^*_{g^{-1}}s({\rm
  Ad_g}\theta),X\rangle$, it follows that, locally, there exists $S(g)$
such that $i_{v^R_X}\Omega_\omega=c d (\langle S(g),X\rangle)
$. Hence, global symmetries are preserved by this deformation provided
$S(g)$ exists globally. This is the case for instance when $H^1(G)=0$
or, as we will see in the next section, when $\omega$ originates from a
group 2-cocycle in $G$. The associated Noether charges are
\begin{equation}
\hat Q_X=-\langle b- c S(g),X\rangle\label{eq:25}. 
\end{equation}
They form a centrally extended
representation of the symmetry algebra,
\begin{equation}
  \label{eq:vcc}
  \pounds_{v^R_{X_1}} \hat Q_{X_2}=\hat Q_{[X_1,X_2]}+c\, \omega(X_1,X_2),
\end{equation}
and are gauge invariant under the modified gauge transformations,
$\pounds_{v^L_{\epsilon}} \hat Q_{X}=0$.

When including a Hamiltonian $\hat H_X=\hat Q_X$, one now finds that
all global symmetries generated by $v^R_{X'}$ with time evolution
determined by \eqref{eq:12} survive if in addition
$c\,\omega(X,X')=0$. When $c\neq 0$, this is a strong condition on
allowed Hamiltonians respecting $G$-invariance.

For a Hamiltonian of the form $\hat H_k=\frac{1}{m!} k^{a_1\dots
  a_m}\hat Q_{a_1}\dots \hat Q_{a_m}$, one has 
\begin{equation}
\label{variationHc}
\pounds_{v^R_{X^{\prime}}} H_k=\frac{1}{(m-1)!} c \, \omega_{bs}
X^{\prime b} \hat Q_{a_2} \cdots
\hat Q_{a_m} k^{sa_2 \cdots a_m } ,
\end{equation}
where $\omega(X,Y)=\omega_{a b} X^a Y^b$. If we restrict ourselves to
field independent Lie algebra elements $X^{\prime}$, 
invariance will hold in the quadratic case, $m=2$, for instance when  
\begin{equation}
\label{evo2c}
\frac{dX^{\prime a}}{dt}= c\, X^{\prime b} \omega_{bs} k^{sa}.
\end{equation}

These deformations will be systematically discussed in the next
section from the viewpoint of centrally extended groups.

\section{Geometric actions for centrally extended groups}
\label{II}

The procedure outlined at the beginning of section \ref{I} can be
straightforwardly generalized for infinite dimensional groups and
central extensions thereof. In applications to three-dimensional
gravity, the asymptotic symmetry algebras of the theory is usually
infinite dimensional, with central extensions in the representation
through surface charges
\cite{Brown:1986nw,Balachandran:1991dw,Banados:1994tn,%
Barnich:2006avcorr}. From the boundary point of view, one should thus
study geometric actions associated to centrally extended groups.

\subsection{Central extensions}
\label{sec:central-extensions}
 
A central extension of a group $G$ is a direct product
$\widehat{G}=G\times\mathbb{R}$, whose elements are pairs
$\left(g,m\right)$ with group operation
$\left(g_1,m_1\right)\left(g_2,m_2\right)=
\left(g_1g_2,m_1+m_2+\Xi(g_1,g_2)\right)$, 
where $\Xi:G\times G\rightarrow\mathbb{R}$ is a $2-$cocycle on $G$
that satisfies
\begin{equation}
  \Xi(g_1g_2,g_3)+\Xi(g_1,g_2)=\Xi(g_1,g_2g_3)
  +\Xi(g_2,g_3),\label{eq:twocc}
\end{equation}
which we assume for simplicitly to satisfy $\Xi(e,g)=0=\Xi(g,e)$. Two
such central extensions denoted by $\Xi$ and $\Xi'$ are isomorphic iff
\begin{equation}
  \label{eq:11}
  \Xi'(g_1,g_2)=\Xi(g_1,g_2)+\mu(g_1)+\mu(g_2)-\mu(g_1g_2),
\end{equation}
where $\mu:G \rightarrow\mathbb{R}$. Denoting the elements of the
corresponding centrally extended Lie algebra
$\widehat{\mathfrak{g}}=\mathfrak{g}\oplus\mathbb{R}$ by $(X,n)$, the
adjoint representation of $\widehat{G}$ can be written as
\begin{equation} {\rm
    Ad}_{\left(g,m\right)}\left(X,n\right)=\left({\rm
      Ad}_{g}X,n-\left\langle S\left(g\right), X\right\rangle
  \right),\label{eq:adgCE}
\end{equation}
where $S:G\rightarrow\mathfrak{g}^{\ast}$ is the Souriau
cocycle on $G$ defined by
\begin{equation}
\label{cocycle}
\left\langle S\left(g\right) ,X \right\rangle = -\left.
  \frac{d}{ds}\left[ \Xi( g,h(s)g^{-1}) +\Xi (
h(s),g^{-1}) \right] \right\vert _{s=0}\text{ },
\end{equation}
with differential at the identity given by
\begin{equation}
  \label{eq:19}
  \frac{d}{ds}S(h(s))|_{s=0}=s(X). 
\end{equation}
Due to \eqref{eq:twocc}, it satisfies the $1-$cocycle condition
\begin{equation}
  S\left(g_1g_2\right)=
  {\rm Ad}^{*}_{{g_2}^{-1}}S\left(g_1\right)+S\left(g_2\right).\label{eq:onecc}
\end{equation}
The adjoint action in $\widehat{\mathfrak{g}}$ is given by 
\begin{equation}
  {\rm ad}_{\left(X,n\right)}\left(Y,k\right)
  =\left({\rm ad}_{X} Y,\omega(X,Y)\right),\label{eq:infadXCE}
\end{equation}
where \eqref{eq:18} has been taken into account and 
where $s$ and ${\rm ad}_{X}$ are the differentials of $S$ and
${\rm Ad}_{g}$ at the identity respectively. Note that $s$ is entirely
determined by the Lie algebra cocycle
$[\omega]\in H^2(\mathfrak g,\mathbb R)$ associated to $\Xi$ according
to equation \eqref{eq:18}. 

Elements in $\widehat{\mathfrak{g}}^{*}$ are denoted by pairs $(b,c)$
where the dual element $c$ to the central extension of $\mathfrak{g}$
is the central charge. The pairing between
$\widehat{\mathfrak{g}}$ and its dual space
$\widehat{\mathfrak{g}}^{*}$ is defined by
\begin{equation}
  \left\langle \left(b,c\right),\left(X,n\right)\right\rangle
  =\left\langle b,X\right\rangle +cn,\label{eq:pairingCE}
\end{equation}
the coadjoint action is given by  
\begin{equation}
  {\rm Ad}_{\left(g,m\right)}^{\ast}\left(b,c\right)=
  \left({\rm Ad}_{g}^{\ast}b-c S\left(g^{-1}\right),c\right),\label{eq:ad*gCE}
\end{equation}
while its the associated action in $\mathfrak g^*$ reads
\begin{equation}
\label{eq:ad*ch}
{\rm ad}_{\left(X,n\right)}^{\ast}\left(b,c\right)
=\left({\rm ad}_{X}^{\ast}b+c s\left(X\right),0\right).
\end{equation}
The extended Maurer-Cartan one-form is denoted by
$(\theta,\theta_{\Xi})$. The additonal piece is  
\begin{equation}
  \label{eq:27}
  \theta_\Xi=dm+\left[\delta_2\Xi(g_1,g_2)\right]\left|_{g_1=g^{-1},g_2=g,\delta
  g_2=dg}\right.,
  \end{equation}
  where $\delta_2$ denotes an infinitesimal variation of $g_2$.
  Equation~\eqref{eq:mceq} is supplemented by
\begin{equation}
d\theta_{\Xi}=\dfrac{1}{2}
  \left\langle s\left(\theta\right),\theta\right\rangle. 
\label{eq:mcCE}
\end{equation}
Differentiating \eqref{eq:onecc} with $g_1=g$ and $g_2=h(s)$ gives
at $s=0$, 
\begin{equation}
\label{diffS}
dS(g)=-{\rm ad}^{\ast}_{\theta} S(g) + s(\theta).
\end{equation}
By using 
\eqref{eq:onecc} applied to $gh(s) g^{-1}$ and differentiating at
$s=0$, with $X$ as in \eqref{eq:adg}, one also gets 
\begin{equation}
  \label{eq:17}
  {\rm ad}^{*}_Y S(g)=-{\rm Ad}^{*}_{g^{-1}} s({\rm Ad}_{g}Y)+s(Y), 
\end{equation}
where $Y={\rm Ad}_{g^{-1}}X$. Combining \eqref{diffS} with
\eqref{eq:17} yields 
\begin{equation}
dS(g)={\rm Ad}^{*}_{g^{-1}}s({\rm
  Ad}_{g}\theta).\label{eq:24}
\end{equation}

\noindent {\bf Remarks:}

(i) Suppose in particular that $H^1(\mathfrak g,\mathbb R)=0$. It can
then be shown that the Souriau map \eqref{eq:18} on the level of the
Lie algebra,
$s: H^2(\mathfrak g,\mathbb R)\to H^1(\mathfrak g,\mathfrak g^*)$,
$[\omega]\mapsto [s]$ is an isomorphism. If furthermore
$H^{2}(\mathfrak g,\mathbb R)$ is of dimension $1$ and the Lie group
$G$ is connected, \eqref{eq:17} determines $S(g)$ uniquely
from $\omega$ (see e.g.~\cite{guieu2007} and original references
therein), without the need for an explicit expression for $\Xi$.

(ii) In the case of a centrally extended group, one can parametrize
the elements of a coadjoint orbit by
$(b,c)={\rm Ad}^{*}_{(g,m)^{-1}}(b_0,c)$. For later use, note that, if
$c\neq 0$ and
\begin{equation}
b_0/c=-S(\Upsilon),\label{eq:15}
\end{equation}
for some group element $\Upsilon$,
it follows from \eqref{eq:onecc} that the coadjoint orbit generated
from $(b_0,c)$ can also be generated from $(0,c)$ provided one changes
$S(g)$ to $S(\Upsilon g)$ in the coadjoint action,
\begin{equation}
  \label{eq:14}
  {\rm Ad}^*_{{(g,m)}^{-1}}(b_0,c)=(-cS(\Upsilon g),c).
\end{equation}

\subsection{Geometric actions for central extensions}
\label{sec:geometric-actions}

The pre-symplectic potential for centrally extended groups is 
$a=\langle(b,c),(\theta,\theta_{\Xi})\rangle$ and the kinetic term
of the geometric action associated to a coadjoint orbit
$O_{(b_{0},c)}$ can be written as
\begin{equation}
  I_{\widehat{\rm G}}[g,m;b_0,c]=I_{\rm G}[g;b_0]+c\int
  \left(-\left\langle S\left(g\right),\theta\right\rangle 
    + \theta_{\Xi}\right). \label{eq:gaCE}
\end{equation}
Using the relations of the previous section, the pre-symplectic 2-form
$\hat\Omega$ for
centrally extended groups can then be worked out
to be $\hat \Omega=\Omega+\Omega_\omega$.

As compared to the analysis at the end of the previous section, the
Lie algebra associated to the centrally extended group has an
additional dimension consisting of vectors of the form $(0,n)$. The
associated left invariant vector fields $v^R_{(0,n)}$ are global
symmetries that generate constant shifts of $m$. They are all trivial
however since these vectors belong to the extended little
algebra. This can also be seen from the fact that
\begin{equation}
  I_{\widehat{\rm G}}[g,m;b_0,c]=I_{\widehat{\rm G}}
  [g,0;b_0,c],\label{eq:29}
\end{equation}
since the dependence on $m$ is only through a total time derivative
that can be omitted. In the following we will simplify the notation
and use $I_{\widehat{\rm G}}[g;b_0,c]$. The additional Noether
charges are trivial constants. More generally, the Noether charges can
be choosen as
\begin{equation}
Q_{(X,n)}=-i_{v^R_{(X,n)}}\langle {\rm Ad}^*_{(g,m)^{-1}}
  (b_0,c),(\theta,\theta_\Xi)\rangle=\hat Q_X-cn,\label{eq:26}
\end{equation}
and now form an ordinary representation of the centrally extended
symmetry algebra, 
\begin{equation}
  \label{eq:vcce}
  \pounds_{v^R_{(X_1,n_1)}} Q_{(X_2,n_2)}=Q_{[(X_1,n_1),(X_2,n_2)]}.
\end{equation}

For orbits generated by $(b_0,c)$ with $c\neq 0$ and where
\eqref{eq:15} holds, it follows from \eqref{eq:14} and the left
invariance of the Maurer-Cartan form that $I_{\rm G}[g;b_0]$ can be
absorbed into the term proportional to the central charge $c$, using
 a new a group element $u=\Upsilon g$, 
\begin{equation} I_{\widehat{\rm G}}[g;b_0,c]=c\int
  \left(-\left\langle S\left(u\right),\theta\right\rangle 
    + \theta_{\Xi}\right)=I_{\widehat{\rm G}}[u;0,c]. \label{eq:gaCE2}
\end{equation}
and analogously, the charges \eqref{eq:25} can be written as 
\begin{equation}
\label{eq:Qr}
\hat{Q}_X=\langle cS(u), X \rangle. 
\end{equation}
This allows one to absorb the term proportional to the orbit
representative $b_0$ also in geometric actions deformed by a
Hamiltonian and to study the geometric actions corresponding to
various coadjoint orbits in a unified fashion.

\subsection{Examples}

As a preparation for the cases of direct interest below, we briefly
revisit in this subsection the well-known geometric actions for
semi-simple loop groups $G$ and for the Virasoro group, first derived
in \cite{Alekseev:1988ce,Rai:1989js}. More details can be found for
instance in \cite{pressley:1986,Khesin2009}. 

\subsubsection{Kac-Moody groups} \label{kmgroup}

\subsubsection*{Loop groups and their extension}

Consider a finite dimensional simple and simply connected group
$G$. The Kac-Moody group $\widehat{{\rm L}G}$ is given by the central
extension of the loop group ${\rm L}G$ of $G$, whose elements
are given by the continuous maps from the unit circle to $G$
\begin{equation}
\label{kmel} g:S^1\to G, \quad \vp \mapsto g(\vp),
\end{equation}
with $g(\vp+2\pi)=g(\vp)$. In the same way, the loop algebra
$L\mathfrak{g}$ corresponds to the algebra of continuous maps from
$S^1$ to $\mathfrak{g}$. The pairing between ${\rm L}\mathfrak{g}$ and
its dual ${\rm L}\mathfrak{g}^{\ast}$ reads
\begin{equation}
  \left\langle
    b\left(\varphi\right),X\left(\varphi\right)\right\rangle
=\int_{0}^{2\pi} d\varphi\  {\rm Tr}\left[ b(\vp) X(\vp)
  \right],  \label{eq:pairingKM} 
\end{equation}
where ${\rm Tr}$ denotes the normalized Killing form. 
The central extension is determined by the $2-$cocycle on the loop
group
\begin{equation}
  \Xi\left(g_1,g_2\right)=\frac{1}{4\pi}\int_{\bar D}{\rm Tr}\left[
  g_1^{-1}\bar{d}g_1\bar{d}g_2g_2^{-1}\right],\label{eq:28}
\end{equation}
where $\bar{d}$ denotes the exterior derivative on the disk $\bar D$
whose boundary is $S^{1}$. The $1-$cocycle defining the adjoint action
can then be obtained from \eqref{cocycle}, 
\begin{equation}
S\left(g\right)=\dfrac{1}{2\pi}g^{-1}\partial_{\varphi}g,\label{eq:SKM}
\end{equation}
with $s\left(X\right)=\frac{1}{2\pi}\partial_{\varphi}X$.  Equation
\eqref{eq:27} gives
\begin{equation}
  \theta_{\Xi}=dm(\vp)+\frac{1}{4\pi}\left(\int_{0}^{2\pi}
    d\varphi\mathrm{Tr}\left[g^{-1}\partial_{\varphi}gg^{-1}dg\right]
    +\int_{\bar{D}}\mathrm{Tr}
    \left[g^{-1}\bar{d}gg^{-1}\bar{d}gg^{-1}dg\right]\right).\label{eq:kmKM}
\end{equation}

\subsubsection*{Geometric actions}

The geometric action (\ref{eq:gaCE}) therefore turns out to be 
\begin{equation}
I_{\widehat{\rm
    LG}}[g;b_0,c]=\int\int_{0}^{2\pi}d\varphi\mathrm{Tr}\left[b_{0}dgg^{-1}
  -\frac{c}{4\pi}g^{-1}\partial_{\varphi}g g^{-1}dg\right] + c\,
\Gamma. \label{eq:gaKM} 
\end{equation}
where
\begin{equation}
\Gamma=\frac{1}{4\pi}\int
\int_{\bar{D}}\mathrm{Tr}\left[g^{-1}\bar{d}gg^{-1}
  \bar{d}gg^{-1}dg\right].\label{eq:gamma} 
\end{equation}
Using the notation $d=dt\partial_{t}$ and defining a manifold
$M=\gamma\times \bar D$ where $t$ is the coordinate along $\gamma$,
the Wess-Zumino term $\Gamma$ can be put into the standard form 
\begin{equation}
  \Gamma= \frac{1}{12\pi}\int_{M}\mathrm{Tr}
  \left[\left(d^Tg g^{-1}\right)^3\right],\label{eq:gamma2}
\end{equation}
where $d^T$ denotes the exterior derivative on the whole of $M$,
$dt \wedge d\varphi\wedge dr$ is considered as orientation for the
integration on $M$, with boundary conditions such that the only
contribution from $\partial M$ arise from $\gamma\times S^1$. 

According to \eqref{eq:26}, the Noether charges associated to the
symmetries corresponding to right multiplication by group elements
$(g(\vp),m(\vp))$ and generated by $(X(\vp),n(\vp))$ are 
\begin{equation}
  Q_{(X,n)}=\int_{0}^{2\pi}d\varphi \,  \left({\rm Tr}\left[Q(\vp)X(\vp)
    \right]-cn(\vp)\right), \quad
  Q(\vp)=\frac{c}{2\pi} g^{-1} \d_\varphi g-g^{-1} b_0
  g. \label{eq:symcurrent} 
\end{equation}
In order to make contact with 3d gravity, we will choose the following
bilinear combination of $Q$ as a Hamiltonian, 
\begin{equation}
\label{HKM}
H_2= \frac{\pi}{c}\int d\vp {\rm Tr} \left[Q^2\right].
\end{equation}
Indeed, this Hamiltonian arises from the Chern-Simons formulation of
AdS$_3$ gravity when imposing Brown-Henneaux boundary conditions
\cite{Coussaert:1995zp}. Under global symmetries generated by
$v^R_{X'}$, it transforms as
\begin{equation}
\label{eq:trans}
\pounds_{v^R_{X'}}H_2=\int d\vp {\rm Tr} \left[Q \d_{\vp} X'
\right].
\end{equation}
Since under the same transformation 
$\delta_{X'} I_{\widehat{\rm LG}}=\int dt d\vp {\rm Tr} \left[Q
  \d_{t} X' \right]$, it follows that
\begin{equation}
\label{eq:tot}
I_{\widehat{\rm LG}}[g;b_0,c,H_2] =I_{\widehat{\rm LG}}[g;b_0,c] -\int dt H_2
\end{equation}
is invariant under the global symmetries generated by $X'$ if
$X'(t,\vp)=X'(t+\vp)$.

\subsubsection*{Relation to chiral WZW theories}

Defining now 
$2\partial_{-}=\d_t-\d_\varphi$, we can write
\begin{equation}
I_{\widehat{\rm LG}}[g;b_0,c,H_2]=-\int dt d\vp \mathrm{Tr}\left[2
 b_{0}g^{-1}\partial_{-}g\right] + I_{\rm
  WZW}[g;c], \label{eq:gawzw2}
\end{equation}
where $I_{\rm WZW}[g;c]$ corresponds to the chiral WZW model
\begin{equation}
I_{\rm WZW}[g;c]=\frac{c}{2\pi}\int dt d\varphi {\rm Tr} \left[
  g^{-1}\partial_{\vp} g g^{-1}\partial_{-}g \right] + c\,
\Gamma, \label{eq:cwzw}
\end{equation}
after neglecting a time independent ${\rm Tr}\, b_0^2$ in the
integrand. In the particular case of a $\vp$ independent $b_0$, action
\eqref{eq:gawzw2} has been obtained in \cite{Elitzur:1989nr} after
solving the constraints of Chern-Simons theory based on a semisimple
group $G$ on a spatial disk with a source. In the context of AdS$_3$
gravity it has been used in \cite{Troost:2003ge} and more recently in
\cite{Kim:2015qoa}.

The term proportional to $b_0$ can be absorbed into the chiral WZW
model by considering a group element $\Upsilon=\Upsilon(\varphi)$ that
solves equation \eqref{eq:15}, which in this case takes the form
\begin{equation}
  \Upsilon^{-1}\d_{\vp}\Upsilon=-\frac{2\pi}{c}b_{0}\,.
\end{equation}
The action
\eqref{eq:gaKM} can then be written as a chiral WZW action for a non
periodic field $u=\Upsilon g$, i.e.,
\begin{equation}
I_{\widehat{\rm LG}}[g;b_0,c,H_2]=I_{\rm WZW}[u,c].
\end{equation}
In this formulation, the dependence on the orbit representative $b_0$
is translated into a nontrivial periodicity of the field $u$,
\begin{equation}
\label{holonomies}
u(\vp+2\pi)=\cM(b_0)u(\vp), \quad \cM(b_0)= \cP
\exp\left[-\frac{2\pi}{c}\oint  d\vp\, b_0\right]. 
\end{equation}

\subsubsection{Virasoro Group}
\label{Virasorogeom}

\subsubsection*{Diffeomorphism group and its extension}

The Virasoro group is the central extension of ${\rm Diff}\left(S^{1}\right)$, which in turn corresponds to
the orientation-preserving diffeomorphism group of the circle with elements $f$ satisfying
\begin{equation}
\label{diffp}
f(\vp+2\pi)=f(\vp)+2\pi, \quad f' >0. 
\end{equation}
The associated Lie algebra will be denoted by
${\rm Vec}\left(S^{1}\right)$. Its elements are vector fields on the
circle $X=X\left(\vp\right)\partial_{\vp}$, while elements of the dual
${\rm Vec}\left(S^{1}\right)^{\ast}$ are taken as quadratic
differentials on $S^{1}$,
$b=b\left(\varphi\right)\left(d\varphi\right)^{2}$.  The natural
pairing between ${\rm Vec}\left(S^{1}\right)$ and its dual is 
\begin{equation}
  \left\langle b,X\right\rangle=\int_{0}^{2\pi}d\varphi \,
b\left(\varphi \right) X\left(\varphi\right).
\label{eq:parV}
\end{equation}
The adjoint and coadjoint actions of ${\rm Diff}\left(S^{1}\right)$
are
\begin{equation} \label{adVir}
  {\rm Ad}_{f^{-1}}\,X = \frac{1}{f' \! \left(\varphi\right)}X
  \left(f \left(\varphi\right)\right)\partial_{\varphi}\, , \quad
{\rm Ad}_{f^{-1}}^{\ast}\,b =  {f^{\prime} \!
  \left(\varphi\right)}^{2}
  \; b\left(f\left(\varphi\right)\right)\left(d\varphi\right)^{2},
\end{equation}
The associated infinitesimal adjoint action is minus the Lie bracket
for vector fields on $S^{1}$.

The 2-cocycle determining the Virasoro group is the Thurston-Bott
cocycle
\begin{equation}
  \Xi(f_1,f_2)=-\frac{1}{48\pi} \int^{2\pi}_0 \, d\vp \log(\d_{\vp}f_1\circ f_2)
  \d_{\vp} (\log(\d_{\vp}f_2)).
\label{TBcocy}
\end{equation}
One then finds
the Schwarzian derivative \eqref{eq:5} as the corresponding Souriau
cocycle, with differential at the identity given by 
\begin{equation}
  s\left(X\right)=\frac{1}{24\pi}
  X^{\prime\prime\prime}\!\left(\vp\right),\label{eq:22}
\end{equation}
while the Maurer-Cartan form is 
\begin{equation}
  \left(\theta,\theta_{\Xi}\right)=\left(\frac{df}{f^{\prime}}\partial_\vp,dm+
    \frac{1}{48\pi}\int_{0}^{2\pi}d\varphi \frac{df}{f^{\prime}}
    \left(\frac{f^{\prime\prime}}{f^{\prime}} \right)' \right).\label{eq:kmVir}
\end{equation}

\subsubsection*{Geometric actions}

The geometric action (\ref{eq:gaCE}) for the Virasoro
group is then found to be 
\begin{eqnarray}
\label{Geometric0}
  I_{\widehat{{\rm Diff}}(S^1)}[f;b_0,c]=\int d\vp \, dt
  \left[b_0(f) f' \dot{f} +\frac{c}{48\pi} \frac{\dot{f}''}{f'} \right] .
\end{eqnarray}
The next step is to add a Hamiltonian preserving diffeomorphisms on
the circle. Again, in order to make contact with three-dimensional
gravity, we chose an Hamiltonian associated with a suitable vector
field. From the discussion of section \ref{I} after equation
\eqref{eq:vcc}, it follows that we may choose $X =-\d_\vp$ with
associated Noether charge
\begin{eqnarray}
\label{Hamiltonian}
H_1=\int d\vp \left[b_0(f) f'^2 +\frac{c}{48\pi}\frac{f''^2}{f'^2} \right]. 
\end{eqnarray}
The invariance under diffeomorphisms
$\delta_{X'} f=X^{\prime} \partial_{\vp} f$ of \eqref{Geometric0} 
then survives the deformation $H_1dt$ if $X^{\prime}$ evolves
according to \eqref{eq:12}. This becomes explicitly
$\d_{t} X^{\prime}={\rm ad}_{-\d_{\vp}} X^{\prime}=\d_{\vp} X^{\prime}$, and implies
$X^{\prime}=X^{\prime}(t+\vp)$.  At this point, it is convenient to
define $\d_{\vp}f=e^\phi$ with $\phi(\vp+2\pi)=\phi(\vp)$. This can be done
because of \eqref{diffp}. In that case, after adding the Hamiltonian
$H_1$ to \eqref{Geometric0}, we find
\begin{eqnarray}
\label{Geometric}
  I_{\widehat{{\rm Diff}}(S^1)} [f;b_0,c,H_1]=  2 \int  d\vp \, dt
  \left[b_0(f) f' \d_- f + \frac{c}{48\pi} \phi' \d_- \phi \right]. 
\end{eqnarray}

\subsubsection*{Relation to chiral bosons}

One may again eliminate the term proportional to the representative
$b_0(\vp)$ from the action by defining a new field with a suitable
periodicity. Following section \ref{II}, this can be done by defining
a new field $F=\Upsilon\circ f$ where $\Upsilon$ satisfies
\begin{equation}
  c\,S_{\varphi}[\Upsilon]=-b_{0}(\varphi).
\end{equation}
The ansatz $F=e^{\mu(f)}$ turns this equation into 
\begin{equation}
  \frac{c}{48\pi}\left(\frac{d\mu}{d f}
  \right)^2-c\,S_{f}[\mu]=b_0(f).\label{eqb}
\end{equation}
In terms of the new field $F$, the action \eqref{Geometric} reduces to
$I_{\widehat{{\rm Diff}}(S^1)}[(F,0);(0,c)]$, which can be rewritten as the
action of a chiral boson $\chi$
\begin{equation}\label{ChBnonperio}
  I_{\rm cB}[\chi;c] =\frac{c}{24\pi} \int dt \,
  d\vp \,\d_{\vp} \chi  \d_-\chi \,,
\end{equation}
where $\d_{\vp}F=e^{\chi}$. In this case, the field redefinition that relates
\eqref{ChBnonperio} with \eqref{Geometric} is
\begin{equation}
\label{eqchi}
\chi =\mu(f) +\phi+ \log \left(\frac{ d\mu}{df} \right) .
\end{equation}
As a side remark, note that \eqref{eqb} turns out to be Hill's
equation: defining $\psi(f)= (\d_{\vp}f)^{1/2}\exp(-\chi/2)$, \eqref{eqb} becomes
\begin{equation}
 \left(-\frac{c}{12\pi}\d^2_{f}
  + b_0(f)\right)\psi(f)=0.
\label{Hill}
\end{equation}

It is well-known that conjugacy classes of monodromy matrices
associated to the Hill's equation characterize Virasoro coadjoint orbits
(see e.g.~\cite{Balog:1997zz}). 

\section{Geometric action for semi-direct products}

\subsection{Semi-direct product groups}

A semidirect product of a semi-simple Lie group $G$ and an abelian
group $\cA$, under some representation $\sigma$ of $G$ on
$\cA$, 
\begin{equation}
\mathcal{S}_\sigma=G\ltimes_{\sigma}\mathcal{A},\label{eq:35}
\end{equation}
is a group with elements of the form $\left(g,\alpha\right)$, where
$g\in G$ and $\alpha \in A$. The group operation is given by
$ \left(g_1,\alpha_1\right) \left(g_2,\alpha_2\right)=\left(g_1 g_2
  ,\alpha_1+\sigma_{g_1}\alpha_2 \right).  $ As $\mathcal{A}$ is
abelian, its Lie algebra is isomorphic to itself and therefore the Lie
algebra associated to $\mathcal{S}_\sigma$ is given by
$\mathfrak{s}=\mathfrak{g}\oright\mathcal{A}$. Denoting the elements
of $\mathfrak{s}$ by $\left(X,\alpha\right)$, and the elements of its
dual space by $\left(j,p\right)$, the bilinear form on $\mathfrak{s}$
is
\begin{equation}
\label{pairsemi} \left\langle
  \left(j,p\right),\left(X,\alpha\right)\right\rangle =\left\langle
  j,X\right\rangle+\left\langle p,\alpha\right\rangle
_{\mathcal{A}}\, ,
\end{equation}
where $\left\langle ,\right\rangle$ and
$\left\langle ,\right\rangle _{\mathcal{A}}$ are the natural pairings
in $\mathfrak{g}$ and $\mathcal{A}$ respectively. The adjoint and
coadjoint actions of $\cS$ on $\mathfrak{s}$ and $\mathfrak{s}^{*}$
\cite{Baguis:1997yp} follow from (\ref{eq:adg}) and (\ref{eq:ad*g}),
\begin{eqnarray} {\rm Ad}_{\left(g,\alpha\right)}\left(X,\beta\right)
& = & \left({\rm Ad}_{g}X,\sigma_{g}\beta -\Sigma_{{\rm
Ad}_{g}X}\alpha\right)\, ,\label{eq:adgSP}\\ {\rm
Ad}_{\left(g,\alpha\right)}^{\ast}\left(j,p\right) & = & \left({\rm
Ad}_{h}^{\ast}j+\sigma_{g}^{\ast}p\mathcal{\odot\alpha},
\sigma_{g}^{\ast}p\right)\, ,\label{eq:ad*gSP}
\end{eqnarray}
where $\Sigma$ is the infinitesimal form of $\sigma$, $p \odot\alpha$
is defined as 
\[
  \left\langle p \odot \alpha,X\right\rangle 
  =\left\langle p,\Sigma_{X}\alpha\right\rangle _{\mathcal{A}}
  =-\left\langle \Sigma_{X}^{\ast}p,\alpha\right\rangle _{\mathcal{A}}
\]
and $\sigma^{\ast},\Sigma^{\ast}$ are the dual maps of $\sigma,\Sigma$
respectively (with respect to the pairing
$\left\langle ,\right\rangle _{\mathcal{A}}$).  The commutation
relations for $\mathfrak{s}$ are defined by the infinitesimal form of
(\ref{eq:adgSP}), i.e.,
\begin{equation}
  \left[\left(X,\alpha\right),\left(Y,\beta\right)\right]
  ={\rm ad}_{\left(X,\alpha\right)}\left(Y,\beta\right)
  =\left({\rm ad}_X Y,\Sigma_{X}\beta
    -\Sigma_{Y}\alpha\right)\,.\label{eq:infadXSP}
\end{equation}

We will construct geometric actions for the group $\cS_\sigma$ when
$\sigma$ corresponds to the adjoint representation and $\mathcal{A}$
is given by the Lie algebra of $G$ seen as an abelian vector space,
which will be denoted by $\mathfrak{g}_{ab}$. Using \eqref{eq:adgSP},
the adjoint action of
$\mathcal{S}_{\rm Ad}=G\ltimes_{{\rm Ad}}\mathfrak{g}_{ab}$ takes the
form
\begin{equation}
  {\rm Ad}_{\left(g,\alpha\right)}\left(Y,\beta\right)
  =\left({\rm Ad}_{g}Y,{\rm Ad}_{g}\beta
    -{\rm ad}_{{\rm Ad}_{g}Y}\alpha\right),\label{eq:adSP2}
\end{equation}
while its infinitesimal form becomes
\begin{equation} \label{adsemi}
  {\rm ad}_{\left(X,\alpha\right)}\left(Y,\beta\right)
  =\left({\rm ad}_{X}Y,{\rm ad}_{X}\beta-{\rm ad}_{Y}\alpha\right).
\end{equation}
For the coadjoint action, (\ref{eq:ad*gSP}) leads to
\cite{Barnich:2014kra,Barnich:2015uva}
\begin{equation}
  {\rm Ad}_{\left(g,\alpha\right)^{-1}}^{\ast}\left(j,p\right)
  =\left({\rm Ad}_{g^{-1}}^{\ast}j-{\rm Ad}_{g^{-1}}^{\ast}
    {\rm ad}_{\alpha}^{\ast} p,{\rm Ad}_{g^{-1}}^{\ast}p\right).\label{eq:ad*SP2}
\end{equation}
The Maurer-Cartan one-form consists of a pair
$\left(\theta,\theta_\alpha\right)$, where $\theta$ is defined in \eqref{eq:mceq} and
\begin{equation}
\label{sa}
\theta_\alpha={\rm Ad}_{g^{-1}} d \alpha.
\end{equation}

\subsection{Geometric actions for $\cS_{\rm Ad}$}
\label{adjointsemipr}

The geometric action (\ref{eq:ga}) for a semi-direct product with an
adjoint action is given by
\begin{equation}
I_{\cS_{\rm Ad}}[g,\alpha;p_0,j_0]=I_{\rm G}[g;j_0]-I_{\rm G}[g;{\rm
  ad}_\alpha^{\ast}p_0].
\label{eq:gaSP}
\end{equation}
In terms of $(h_L,\alpha_L)$ and $(h_R,\alpha_R)$, left and right
actions in a semi-direct product group act as
\begin{equation}
 \label{eq:sdmultI}
\begin{split}
g\to h_L  g,& \quad \alpha \to \alpha_L + {\rm Ad}_{h_L} \alpha,\\
g\to g  h_R,& \quad \alpha \to \alpha+ {\rm Ad}_{g} \alpha_R.
\end{split}
\end{equation}
Geometric actions \eqref{eq:gaSP}  are invariant under gauge and global transformations generated by vector fields  $v^R_{(X,n)}$ and $v^L_{(\epsilon,\zeta)}$ respectively. They are given by
\be
\label{vectoresemi}
\begin{split}
v^R_{(X,\upsilon)}&=\frac{d}{ds}(g h_R(s) , \alpha+ {\rm Ad}_{g} \alpha_R(s))|_{s=0}, \, X=\frac{d}{ds} h_R(s)|_{s=0}, \, \upsilon=\frac{d}{ds} \alpha_R(s)|_{s=0},\\
v^L_{(\epsilon,\zeta)}&=\frac{d}{ds}( h_L(s)  g , \alpha_L(s) + {\rm Ad}_{h_L(s)} \alpha )|_{s=0}, \, \epsilon=\frac{d}{ds} h_L(s)|_{s=0}, \, \zeta=\frac{d}{ds} \alpha_L(s)|_{s=0}.
\end{split}
\ee
where  $(\epsilon, \zeta)$ depends arbitrary on time $t$ and belongs to the little algebra of the representatives $(p_0,j_0)$.

\subsection{Geometric action for centrally extended
  $\widehat{\cS}_{\rm Ad}$}

Let us consider now the centrally extended group
$\widehat{\mathcal{S}}=\widehat{G}\ltimes_{{\rm
    Ad}}\widehat{\mathfrak{g}}_{ab}$, i.e. a central extension of a
semi-direct product group under the adjoint action, whose elements
will be denoted by $\left(g,m_1,\alpha,m_2\right)$. The elements of
the algebra $\widehat{\mathfrak{s}}$ will be denoted by
$\left(X,n_1,\alpha,n_{2}\right)$, while the elements of the dual by
$\left(j,c_{1},p,c_{2}\right)$.  The coadjoint action as well as for
the geometric action are constructed using (\ref{eq:ad*SP2}) and
(\ref{eq:gaSP}) and replacing
\begin{eqnarray*}
j_{0} \to  (j_{0},c_{1}),\quad 
p_{0} \rightarrow  (p_{0},c_2),\quad 
{\rm Ad}_{g}^{*} \rightarrow  {\rm Ad}_{(g,m_{1})}^{*},\quad
{\rm ad}_{\alpha}^{\ast} \rightarrow  {\rm ad}_{(\alpha,n_{2})}^{*},
\end{eqnarray*}
where ${\rm Ad}_{(g,m_{1})}^{*}$ is defined in (\ref{eq:ad*gCE}) and
${\rm ad}_{(\alpha,n_{2})}^{*}$ corresponds to its infinitesimal form
given by \eqref{eq:ad*ch}.  In the same way, the one-form
$(\theta,\theta_\alpha)$ introduced in the previous section must be
replaced by
$\left(\theta,\theta_\Xi,\theta_\alpha,\theta_\omega\right)$ where
$ \theta_\omega=\langle S(g), \theta_\alpha
  \rangle$.
  
The  geometric action on a coadjoint orbit
$O_{\left(j_{0},c_{1},p_{0},c_{2}\right)}$ is
given by the centrally extended version of \eqref{eq:gaSP}, 
\begin{equation}
I_{\widehat{\cS}_{\rm Ad}}[g,\alpha;p_0,j_0,c_1,c_2] = I_{\widehat{\rm
    G}}[g;j_0,c_1]
-I_{\rm G}[g;{\rm ad}_\alpha^{\ast}p_0+c_2 s(\alpha)],
\label{eq:gaCESP}
\end{equation}
where $I_{\widehat{\rm G}}$ is given by \eqref{eq:gaCE}. 

As before, the orbit representatives $j_{0}$ and $p_{0}$ can be absorbed
into the terms proportional to the central charges by defining
suitable fields $u=\Upsilon g$ and
$a=\eta+{\rm Ad}_{\Upsilon}\alpha$. Using \eqref{eq:17} allows one
to write the action \eqref{eq:gaCESP} in the form
\begin{multline}
I_{\widehat{\cS}_{\rm Ad}}[g,\alpha;p_0,j_0,c_1,c_2] = I_{{\mathcal
    S}}[g,\alpha;j_{0},p_{0}]
+c_1\int \langle {\rm Ad}^{*}_{g^{-1}}S(\Upsilon),\theta\rangle \\
+c_2\int
\langle {\rm Ad}^{*}_{g^{-1}}\left({\rm
    Ad}^{*}_{\Upsilon^{-1}}s(\eta)
  -{\rm ad}^{*}_{\alpha}S(\Upsilon)\right),\theta\rangle
+I_{\widehat{\cS}_{\rm Ad}}[u,a;0,0,c_1,c_2] .\label{eq:gaCESP2}
\end{multline}
An inspection of the latter expression makes evident that, provided
the pair $(\Upsilon,\eta)$ satisfies
\begin{equation}
  c_2
S(\Upsilon)=-p_0,\quad c_2 {\rm
  Ad}^{*}_{\Upsilon^{-1}}s(\eta)=-j_0+\frac{c_1}{c_2}p_0
\,,\label{eq:rep}
\end{equation}
the geometric action reduces to
\begin{equation}
I_{\widehat{\cS}_{\rm Ad}}[g,\alpha;p_0,j_0,c_1,c_2] =
I_{\widehat{\cS}_{\rm Ad}}[u,a;0,0,c_1,c_2] .\label{eq:gaCESP3}
\end{equation}
As we will
see in the next examples, the latter equality will allow us to link
geometric actions based on groups having a semi-direct product
structure with actions appearing in the Hamiltonian reduction of 3d
gravity in the case of vanishing cosmological constant.

\subsection{Examples}

\subsubsection{Loop groop of $G\ltimes\mathfrak{g}$ and its extension}

Let us consider the group
$\hat{\cS}_{\rm Ad}=\widehat{L{\rm G}}\ltimes_{\rm Ad}\widehat{{\rm
    L}\mathfrak{g}}_{\rm ab}$, where $G$ is a semi-simple Lie
group. Its elements are of the form $(g(\vp),m_1,\alpha(\vp),m_2)$,
where $g(\vp)$ is a map of the form \eqref{kmel}, $\alpha$ is an
element the loop algebra ${\rm L}\mathfrak{g}$ and $m_1$, $m_2$
correspond to the central extensions of ${\rm L}G$ and
${\rm L}\mathfrak{g}$ respectively.  The geometric action can be
obtained directly from (\ref{eq:gaCESP}) using the machinery developed
in subsection \ref{kmgroup}, 
\begin{multline}
\label{eq:gaSPKM}
I_{\widehat{L{\rm G}}\ltimes_{\rm Ad}\widehat{{\rm
    L}\mathfrak{g}}_{\rm ab}}[g,\alpha;p_0,j_0,c_1,c_2]  =\\
I_{\widehat{\rm LG}}[g;j_0,c_1]-\int d\vp\mathrm{Tr}
\left[ \left([\alpha,p_0]+\frac{c_{2}}{2\pi}\alpha^{\prime}
  \right)dg g^{-1}\right],
\end{multline}
where $I_{\widehat{\rm LG}}$ is given by \eqref{eq:gaKM}. From
\eqref{eq:sdmultI}, this action is invariant under right multiplication
of $g$, but also under the adjoint action on $\alpha$, i.e., under 
\begin{equation}
\label{eq:pl1}
\delta_{(X,\upsilon)}(g,\alpha)=\left(gX(\vp), g \upsilon(\vp) g^{-1} \right).
\end{equation}
These global symmetries give rise to the following Noether charges
\begin{equation} \label{isoQ}
\begin{split}
&\cJ_X= \int^{2\pi}_0 d\vp\, {\rm Tr} \left[X j \right],
\quad j=\frac{c_1}{2\pi}   g^{-1} \d_{\vp} g- g^{-1} \left(j_0
  -\frac{c_2}{2\pi} \d_{\vp} \alpha -[\alpha, p_0]\right)g,\\
&\cP_\upsilon=\int^{2\pi}_0 d\vp\, {\rm Tr} \left[\upsilon p \right],
\quad p=\frac{c_2}{2\pi} g^{-1} \d_{\vp} g-g^{-1} p_0 g.
\end{split}
\end{equation}

A Hamiltonian motivated by asymptotically flat gravity in three
dimensions is 
\begin{equation}
\label{Hsm}
H_2=\frac{\pi}{c_2} \int d\vp {\rm Tr} \left[p^2\right].
\end{equation}
Since $H_2$ transforms as \eqref{eq:trans} under the symmetries
\eqref{eq:pl1} and
$\delta_{(X,\upsilon)}I_{\widehat{L{\rm G}}\ltimes_{\rm Ad}\widehat{{\rm
      L}\mathfrak{g}}_{\rm ab}}$ is given by
$\int dt d \vp {\rm Tr}\left[\d_t \upsilon p + \d_t X j\right]$, one
concludes that in presence of the Hamiltonian, action
\begin{equation}
\label{ISH}
I_{\widehat{L{\rm G}}\ltimes_{\rm Ad}\widehat{{\rm
    L}\mathfrak{g}}_{\rm ab}}[g,\alpha;p_0,j_0,c_1,c_2] - \int dt H_2 
\end{equation}
is invariant provided $X= X_0(\vp)$, $\upsilon= \upsilon_{0}(\vp) + t \d_{\vp}X_0$.
\subsubsection*{Relation to Flat WZW model}

The terms in \eqref{eq:gaSPKM} proportional to the orbit
representatives $j_0$ and $p_0$ can be absorbed into the kinetic term
of the flat WZW model by defining new fields $u=\Upsilon g$ and
$a=\eta+\Upsilon\alpha\Upsilon^{-1}$ satisfying equations
\eqref{eq:rep} , which in this case take the form
\begin{equation}
  -\frac{c_2}{2\pi}\Upsilon^{-1} \d_{\vp} \Upsilon=p_0, \quad
  -\frac{c_2}{2\pi}\Upsilon^{-1}\d_{\vp}\eta\Upsilon
  =j_0-\frac{c_1}{c_2}p_0 \,.  
\end{equation}
After including the Hamiltonian, the geometric action \eqref{ISH} can
be written in terms of the new fields $u$ and $a$ as
\begin{multline}
I_{\widehat{L{\rm G}}\ltimes_{\rm Ad}\widehat{{\rm
    L}\mathfrak{g}}_{\rm ab}}[g,\alpha;p_0,j_0,c_1,c_2]
=\\ I_{\widehat{\rm LG}}[u;0,c_1]-\frac{c_{2}}{2\pi}\int d\vp dt
\mathrm{Tr}\left[\dot{u} u^{-1}a^{\prime}
  -\frac{1}{2} (u^{-1}u^{\prime})^2\right].
\label{eq:gaSPKM2}
\end{multline}
This corresponds to the flat WZW model obtained in
\cite{Barnich:2013yka,Barnich:2015sca} in the context of
asymptotically flat three-dimensional Einstein gravity. In this
representation, the information on $(p_0,j_0)$ is encoded in the
periodicity of the fields $u$ and $a$,
\begin{equation}
u(\vp+2\pi)=\cM(p_0)u(\vp)\,, \quad a(\vp+2\pi)=\cM(p_0) a(\vp)
\cM^{-1}(p_0) + \cN(j_0,p_0),
\end{equation}
where $\cM(p_0)$ is given by \eqref{holonomies} and
$\cN(j_0,p_0)=-\frac{2\pi}{c_2}\oint \Upsilon \left(
  j_0-\frac{c_1}{c_2}p_0 \right) \Upsilon^{-1}$.

\subsubsection{$\widehat{\rm BMS}_{3}$ group}

The $\widehat{\rm BMS}_{3}$ group is the semidirect product of the
Virasoro group and its algebra (seen as an abelian vector space) under
the adjoint action
\[
  \widehat{\rm BMS}_{3}=\widehat{\rm Diff}\left(S^{1}\right)
  \ltimes {\rm Vec}\left(S^{1}\right)_{\rm ab}.
\]
Its elements are pairs $(f,\alpha)$, where $f$ is a diffeomorphism of
the circle \eqref{diffp} and $\alpha$ satisfies
\begin{equation}
\label{alpha}
\alpha(\vp+2\pi)=\alpha(\vp)\,. 
\end{equation}
Therefore, the corresponding geometric action has the form (\ref{eq:gaCESP})
where the coadjoint action is the one of the Virasoro group (\ref{adVir})
and $S$ is the Schwarzian derivative (\ref{eq:5}). The resulting action is
\begin{multline}
\label{GeometricBMS}
I_{\widehat{\rm BMS}_{3}}[f, \alpha; p_0,j_0,c_1,c_2]=\\
I_{\widehat{{\rm Diff}}(S^1)} [f;j_0,c_1]+\int d\vp \left[f^{\prime}df
  \left(p_{0}^{\prime}\alpha+2p_{0}\alpha^{\prime}
    -\frac{c_{2}}{24\pi}\alpha^{\prime\prime\prime}\right)\circ f\right].
\end{multline}
Defining 
$
e^\phi = f^{\prime}$, 
$\xi = \alpha^{\prime}\left(f \right)
$, the geometric
action on a orbit of the $\widehat{\rm BMS}_{3}$ group takes the form
\begin{multline}
  I_{\widehat{\rm BMS}_{3}}[f, \alpha; p_0,j_0,c_1,c_2]
  =\\I_{\widehat{{\rm Diff}}(S^1)} [f;j_0,c_1]+\int d\varphi dt
  \left[f^{\prime}\dot{f} \left(p_{0}^{\prime}\alpha+
      2p_{0}\alpha^{\prime}\right)\circ f
    +\frac{c_{2}}{24\pi}\dot{\phi}\xi^{\prime}\right].\label{eq:ibms}
\end{multline}
From \eqref{eq:sdmultI}, we can infer that the global transformations laws of the fields are
\be
\label{trBMS}
\delta_{(X,\upsilon)} (f, \alpha(f)) = (X(\vp)\d_{\vp} f, \upsilon(\vp) \d_{\vp}f).
\ee
We will choose the Hamiltonian as the charge associated to rigid translation, $(X,\upsilon)=(0,-\d_{\vp})$
\begin{equation}
\label{SchwBMS}
H=\int ^{2 \pi}_0 d\vp \, \left[ f'^2 p_0(f)
    +\frac{c_2}{48\pi}\frac{f''^2}{f'^2} \right].
\end{equation}
 As in the previous examples, this choice is inspired by three-dimensional Einstein gravity without cosmological constant. Let us now consider the geometric action \eqref{eq:ibms} deformed by the Hamiltonian \eqref{SchwBMS}. Symmetry \eqref{trBMS} will be preserved in this new action provided the suitable extension of \eqref{eq:12} is satisfied, i.e.
$\d_t (X', \upsilon')= {\rm ad}_{(0,-\d_{\vp})} (X',\upsilon')
$, which gives $X'=X_0(\vp)$ and $\upsilon'=\upsilon_0(\vp)+t \d_{\vp}X_0$.

\subsubsection*{Relation to chiral BMS$_3$ theory}

Defining new fields $F=\Upsilon \circ f$ and
$a=\eta+\Ad_{\Upsilon} \alpha$ satisfying
\eqref{eq:rep}, which in this case takes the form
\begin{equation}
\label{hillbms}
  c_2\,S_{\varphi}\left[\Upsilon\right]=-p_0, \quad
  -\frac{c_2}{24\pi}\Upsilon^{\prime\,2}
  \eta^{\prime\prime\prime}(\Upsilon)=j_0-\frac{c_1}{c_2}p_0 \,,
\end{equation}
the terms in \eqref{eq:ibms} proportional to the orbit representatives
$j_0$ and $p_0$ can be absorbed in this field redefinition. Including
$H$ in \eqref{eq:ibms}, the geometric action for the ${\rm BMS_{3}}$
group can be written as
\begin{equation}
  I_{\widehat{\rm BMS}_{3}}[f, \alpha; p_0,j_0,c_1,c_2,H] =
  I_{{\widehat{{\rm Diff}}(S^1)}}[\chi;0,c_1]+\frac{c_{2}}{24\pi}
  \int d\varphi dt \,\left(\dot{\chi}\zeta^{\prime}
    -\frac{1}{2}\chi^{\prime2}\right) \,, \label{eq:ibms2}
\end{equation}
where $I_{{\widehat{{\rm Diff}}(S^1)}}[\chi;0]$ is given in
\eqref{ChBnonperio}, $\chi={\rm log}(\d_{\vp} F)$ and
$\zeta=a^{\prime}(F)$. This is the chiral BMS$_3$ model constructed as
the classical dual of three-dimensional asymptotically flat gravity
\cite{Barnich:2013yka,Barnich:2015sca} (See also \cite{Gonzalez:2014tba} 
for a higher spin extension).

Note that, as in the Virasoro case, the first 
relation of \eqref{hillbms} produces the Hill's equation for the variable 
$\psi(f)$ defined in \eqref{Hill} with $(p_0,c_2)$ playing the role of $(b_0,c)$.
In the same way, the second equation of \eqref{hillbms} controls the orbits associated to
the pair $(j_0,c_1)$.

\section{Discussion and perspectives}

We have studied geometric actions for various groups arising in
three-dimensional gravity. In a first stage, we have analyzed
geometric actions for loop groups reobtaining \eqref{eq:gaKM}, and
constructing \eqref{eq:gaSPKM} for the semi-direct product
case. Introducing monodromies in the groups elements, these actions
can be written as chiral WZW models \eqref{eq:gaSPKM2} and
\eqref{eq:cwzw}. When considering Chern-Simons theories on manifolds
with non contractible cycles, it has been shown in
\cite{Elitzur:1989nr} that the term proportional to $b_0$ in
\eqref{eq:gaKM} arises in the associated WZW theory. Thus, one should
expect these models to originate from the Chern-Simons formulation of
gravity after solving the constraints inside the action once
holonomies are properly taken into account. More precisely, this would
be the case when adopting the new boundary conditions that have been
proposed recently in the context of three dimensional gravity
\cite{Grumiller:2016pqb}, \cite{Grumiller:2017sjh} and that give rise
to loop groups as asymptotic symmetry groups: for AdS$_3$ gravity the
charge algebra yields two copies of $\hat{\rm{SL}(2,\mathbb{R})}$,
while in the case of vanishing cosmological constant, it is the
centrally extended Poincar\'e loop group
$\hat{{\rm SL}(2,\mathbb{R}) \ltimes \mathfrak{sl}(2,\mathbb{R})}$
that appears. In this sense, actions \eqref{eq:gaKM} and
\eqref{eq:gaSPKM} are the $(1+1)$-models representing these boundary
degrees of freedom.

In the case of Brown-Henneaux boundary conditions, the asymptotic
symmetry group is given by two copies of the diffeomorphism group of
the circle and the dual dynamics is controlled by the difference of
two ${\rm Diff}(S^1)$ invariant actions \eqref{Geometric}. For
asymptotically flat spacetimes, similar boundary conditions lead to a
boundary dynamics that is controlled by the BMS$_3$ invariant model
\eqref{GeometricBMS}. In both cases, once we remove the representative
term, we obtain the difference of two chiral bosons
\eqref{ChBnonperio} for AdS$_3$ boundary conditions, and
\eqref{eq:ibms2} in the case of flat geometries. These results are
consistent with the earlier derivations
\cite{Henneaux:1999ib,Barnich:2013yka}. Note however that the
periodicity of the chiral fields is determined by the value of the
representative(s).
 
The results of this paper can readily be generalized to other
situations that arise in the context of three-dimensional gravity: one
could for instance use the general formula \eqref{eq:gaCESP} to work
out geometric actions associated to other groups with semi-direct
product structure like Warped Virasoro \cite{Compere:2013bya} or
extensions of BMS$_3$ with spin-one generators
\cite{Detournay:2016sfv}.

Geometric actions can be used to compute one-loop partition functions
associated to three-dimensional gravity. It has been shown in
\cite{Alekseev:1990mp} for the Virasoro case and for constant
representatives that the partition function leads to characters
associated with highest weight representations. The key ingredient in
this derivation was provided by a transformation that removes the
representative from the action. In this paper, we have found a
generalization of that transformation for any centrally extended
group. Thus, we hope to prove more general connections between
characters and partition functions by using relations
\eqref{eq:15} and \eqref{eq:gaCE2}.

Another interesting direction corresponds to exploring the
connection established between geometric actions and Berry phases
\cite{Oblak:2017ect}. More precisely, it would be interesting to
understand the physical content of these phases in the case of systems
with BMS$_3$ symmetry or other groups with a semi-direct product
structure.

Coadjoint orbits have also appeared recently in the study of the SYK
model (see e.g. \cite{Maldacena:2016hyu} for a review). In
\cite{Stanford:2017thb}, the Hamiltonian associated to a rigid
rotation \eqref{Hamiltonian} in the case where $b_0= -\frac{c}{48\pi}$
is taken as the Euclidean action of the model. In this regard, it
would be interesting to understand whether other conserved charges for
the Virasoro group, for BMS$_3$ symmetry, or for
the Poincar\'e loop group, could play a similar
role. 

A most relevant extension of the considerations here consists in
modifying from the very beginning the set-up of section \bref{I} by
letting $b_0=b_0(t)$ be a dynamical variable. This means that one no
longer considers the dynamics on a fixed coadjoint orbit, but rather a
suitable collection of orbits and the associated dynamics. In
practice, this can be done for instance by introducing an additional
vector $a_0=a_0(t)\in\mathfrak g$. One may then choose the extended
pre-symplectic potential
\begin{equation}
  \label{eq:31}
  a^E=a+\langle b_0,da_0\rangle, 
\end{equation}
and the extended kinetic term
\begin{equation}
I^E_{\rm G}[g,b_0,a_0]=\int_{\gamma^E} a^E.\label{eq:30}
\end{equation}
The associated pre-symplectic $2$-form is
\begin{equation}
  \label{eq:32}
  \Omega^E=\Omega+\langle db_0,{\rm
    Ad}_{g}\theta\rangle+\langle b_0,d a_0\rangle.
\end{equation}
The extended equations of motion are equivalent to 
\begin{equation}
i_V\Omega+\langle
\dot b_0,{\rm Ad_g} i_V\theta\rangle=0,\quad \dot a_0=-{\rm Ad}_gi_V
\theta,\quad \dot b_0=0.\label{eq:33}
\end{equation}
Hence, when choosing the integration constants $b_0(t)=\bar b_0$ to
coincide with the constant values of section \bref{I}, the dynamics
of the group variables $g$ is unchanged. The additional integration
constants $\bar a_0$ are controlled by additional global symmetries
that correspond to constant shifts of $a_0$.

In particular, in the case of centrally extended groups, we get
the kinetic term
\begin{equation}
I_{\widehat G}[(g,m),(b_0,c),(a_0,d)].\label{eq:34}
\end{equation}
In this case, the group element $m$ no longer drops out of the problem
and the associated global symmetry becomes relevant, and so does the
quadratic term $b_0^2$ in section \ref{kmgroup}. Both the orbit
representative $b_0$ and the central charge $c$ are now dynamical
variables. For the orbit representative, this has been analyzed from
the geometric actions point of view in the context of ``model spaces''
in \cite{Alekseev:1990mp,La:1990ty} and from the bulk of viewpoint in
\cite{Compere:2015knw}. For the central extension, it seems that such
a generalization has not yet been considered in the context of
geometric actions, whereas from the bulk viewpoint it has recently
been discussed in \cite{Bunster:2014cna}.

Finally, we will discuss elsewehere the implications of the extended
set-up for three-dimensional gravity. In particular, we will study in
more detail $(i)$ the bulk duals of the geometric actions obtained
here in terms of suitable choices of boundary conditions and by
properly taking into account the holonomies and the dynamics of the
associated particles, $(ii)$ the interpretation in terms of Goldstone
bosons and the connection to non-linear realizations. Most
importantly, since the present framework is entirely
group-theoretical, there are a priori no obstructions to constructing
dynamical actions appropriate to BMS symmetry in four dimensions.

\section*{Acknowledgements}

\addcontentsline{toc}{section}{Acknowledgments}

We thank Andr\'es Gomberoff, Daniel Grumiller, Wout Merbis and Blagoje
Oblak for useful discussions. G.B.~is grateful to Fondecyt (Chile)
Grant N$^{\circ}$1141309 for support during his visit to Chile where
part of this work was completed. G.B.~is supported by the Fund for
Scientific Research-FNRS (Belgium) (convention FRFC PDR T.1025.14 and
convention IISN 4.4503.15), H.G.~is supported by the Austrian Science
Fund (FWF), project P 28751-N2, and P. S-R.~is supported by the
Fondecyt (Chile) Grant N$^{\circ}$3160581.


\end{document}